\documentclass[]{aa}
\usepackage{comment}
\usepackage{graphicx}
\usepackage{txfonts}
\usepackage[svgnames]{xcolor}
\usepackage[colorlinks=true, linkcolor = blue, urlcolor =magenta, hyperfootnotes=false]{hyperref}

\usepackage{footmisc}
\newcommand{\astfootnote}[1]{%
\let\oldthefootnote=\thefootnote%
\setcounter{footnote}{0}%
\renewcommand{\thefootnote}{\fnsymbol{footnote}}%
\footnote{#1}%
\let\thefootnote=\oldthefootnote%
}

\usepackage{tablefootnote}
\usepackage{bm}

\hypersetup{
    colorlinks,
    citecolor=blue
}

\usepackage{natbib}
\bibpunct{(}{)}{;}{a}{}{,} 
\usepackage{subfigure}

\begin{document}

   \title{\texttt{KOBEsim}: A Bayesian observing strategy algorithm for planet detection in radial velocity blind-search surveys\thanks{Based on observations collected at Centro Astron\'omico Hispano en Andaluc\'ia (CAHA) at Calar Alto, operated jointly by Instituto de Astrof\'isica de Andaluc\'ia (CSIC) and Junta de Andaluc\'ia.}}

   \author{O.~Balsalobre-Ruza
          \inst{\ref{cab}}
          \and
          J.~Lillo-Box\inst{\ref{cab}}
          \and
          A.~Berihuete\inst{\ref{cadiz}}
          \and
          A.~M.~Silva\inst{\ref{IACE},\ref{porto}}
          \and
          N.~C.~Santos\inst{\ref{IACE},\ref{porto}}
          \and
          A.~Castro-González\inst{\ref{cab}}
          \and
          J.~P.~Faria\inst{\ref{IACE},\ref{porto}}
          \and
          N.~Huélamo\inst{\ref{cab}}
          \and
          D.~Barrado\inst{\ref{cab}}
          \and
          O.~D.~S.~Demangeon\inst{\ref{IACE},\ref{porto}}
          \and
          E.~Marfil\inst{\ref{cab}}
          \and
          J.~Aceituno\inst{\ref{caha},\ref{IAA}}
          \and
          V.~Adibekyan\inst{\ref{IACE},\ref{porto}}
          \and
          M.~Azzaro\inst{\ref{caha}}
          \and
          S.~C.~C.~Barros\inst{\ref{IACE},\ref{porto}}
          \and
          G.~Bergond\inst{\ref{caha}}
          \and
          D.~Galad\'{\i}-Enr\'{\i}quez \inst{\ref{caha}}
          \and
          S.~Pedraz\inst{\ref{caha}}
          \and
          A.~Santerne\inst{\ref{france}}
          }

   \institute{Centro de Astrobiología (CAB), CSIC-INTA, Camino Bajo del Castillo s/n, 28692, Villanueva de la Cañada, Madrid, Spain\\
   e-mail: {\tt obalsalobre@cab.inta-csic.es}\label{cab}
   \and
   Depto. Estadística e Investigación Operativa, Universidad de Cádiz, Avda. República Saharaui s/n, 11510 Puerto Real, Cádiz, Spain\label{cadiz}
   \and
   Instituto de Astrof\'isica e Ci\^encias do Espa\c{c}o, Universidade do Porto, CAUP, Rua das Estrelas, 4150-762, Porto, Portugal\label{IACE}
   \and
   Departamento de F\'isica e Astronomia, Faculdade de Ci\^encias, Universidade do Porto, Rua do Campo Alegre, 4169-007 Porto, Portugal\label{porto}
   \and
   Centro Astron\'omico Hispano en Andaluc\'\i a, Observatorio de Calar Alto, Sierra de los Filabres, 04550 G\'ergal, Almer\'\i a, Spain\label{caha}
   \and
   Instituto de Astrofísica de Andalucia , Glorieta de la Astronomia s/n, Granada, Spain\label{IAA}
   \and
   Aix Marseille Univ, CNRS, CNES, LAM, Marseille, France\label{france}
   }
   \date{Received 3 May 2022 / Accepted 20 October 2022}

  \abstract
  {Ground-based observing time is precious in the era of exoplanet follow-up and characterization, especially in high-precision radial velocity instruments. Blind-search radial velocity surveys thus require a dedicated observational strategy in order to optimize the observing time, which is particularly crucial for the detection of small rocky worlds at large orbital periods.
   }
   {We developed an algorithm with the purpose of improving the efficiency of radial velocity observations in the context of exoplanet searches, and we applied it to the K-dwarfs Orbited By habitable Exoplanets (KOBE) experiment. Our aim is to accelerate exoplanet confirmations or, alternatively, reject false signals as early as possible in order to save telescope time and increase the efficiency of botAh blind-search surveys and follow-up of transiting candidates.}
   {Once a minimum initial number of radial velocity datapoints is reached in such a way that a periodicity starts to emerge according to generalized Lomb-Scargle (GLS) periodograms, that period is targeted with the proposed algorithm, named \texttt{KOBEsim}. The algorithm selects the next observing date that maximizes the Bayesian evidence for this periodicity in comparison with a model with no Keplerian orbits.}
   {By means of simulated data, we proved that the algorithm accelerates the exoplanet detection, needing 29 -- 33\,\% fewer observations and a 41 -- 47\,\% smaller time span of the full dataset for low-mass planets ($m_{\rm p}$ $<$ 10\,$M_{\oplus}$) in comparison with a conventional monotonic cadence strategy. For 20\,$M_{\oplus}$ planets we found a 16\,\% enhancement in the number of datapoints. We also tested \texttt{KOBEsim} with real data for a particular KOBE target and for the confirmed planet \object{HD~102365\,b}. These two tests  demonstrate that the strategy is capable of speeding up the detection by up to a factor of 2 (i.e., reducing both the time span and number of observations by half).}
   {}


   \titlerunning{\texttt{KOBEsim}}
   \maketitle
%

\section{Introduction}

  Over the last three decades numerous exoplanets have been detected (more than 5000 confirmed according to the NASA Exoplanet Archive\footnote{\url{https://exoplanetarchive.ipac.caltech.edu/}}, \citealt{Akeson_2013}). Ever since the first hot Jupiters were discovered (\citealt{1995Natur.378..355M}; \citealt{1997ApJ...474L.115B}) the instrumentation has improved, thus making it possible to detect  less massive worlds. Specifically in the radial velocity (RV) method, the development of highly stabilized spectrographs, from CORALIE (\citealt{2000A&A...354...99Q}) to ESPRESSO (\citealt{2021A&A...645A..96P}), have allowed the detection of lighter planets located farther away from the host star (e.g., \citealt{2020A&A...642A..31D};
  \citealt{2021A&A...653A..41D};
  \citealt{lillobox2021hd22496b};
  \citealt{2022A&A...658A.115F}). However, detection and characterization are still challenging, especially in non-transiting planetary systems since they require many observations distributed over long periods of time. When searching for potentially habitable worlds, this becomes crucial as they orbit at larger periods than those typically detected. That is why the improvement in the efficiency of observations is decisive. It saves valuable telescope time and allows us to find a type of planet that would otherwise be undetectable throughout conventional observing programs.

 \begin{figure*}
  \begin{center}
      \subfigure{\includegraphics[width=180mm]{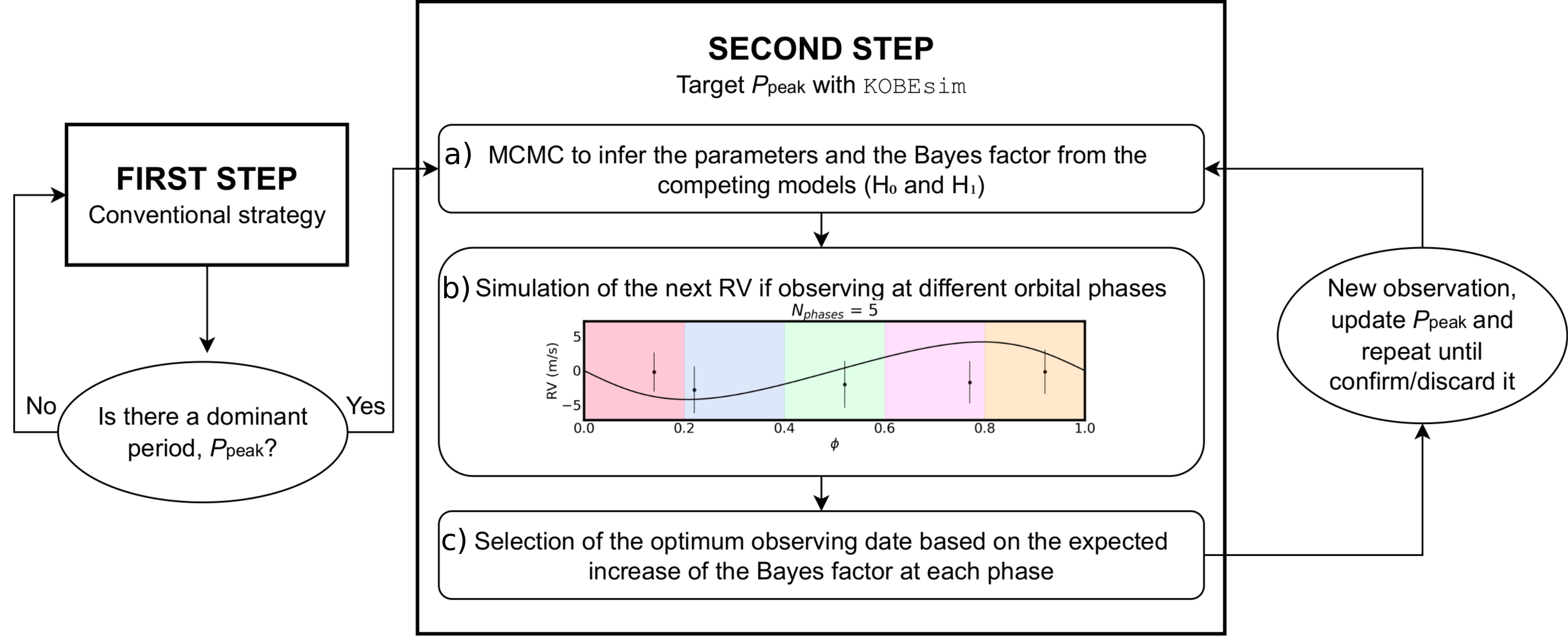}}\vspace{0.25cm}
  \end{center}
 \caption{\texttt{KOBEsim} workflow scheme. It starts by gathering RV data until it finds a dominant period (step 1), from which the  \texttt{KOBEsim} algorithm provides the next optimum observing date (step 2).}
 \label{fig:diagram}
\end{figure*}

  To address the problem of observational efficiency, it is increasingly common that observatories and scientific programs develop their own tools to avoid wasting telescope time through a good scheduling strategy. Some of these approaches have been presented, such as for ALMA (\citealt{2014SPIE.9149E..1SE}), JWST (\citealt{giu}), and the CARMENES Guaranteed Time Observation (GTO) survey (\citealt{2017A&A...604A..87G}). In the context of RV searches, \citet{2021MNRAS.503.5504C} proposed that uniformly distributed data along the phase-folded RV diagram favors efficiency. As a consequence, observing in an already explored orbital phase ($\phi$) does not significantly increase the information in hand. In this paper we present \texttt{KOBEsim}\footnote{The code is publicly available at the following link: \url{https://github.com/olgabalsa/KOBEsim}}, a Bayesian algorithm to improve the efficiency of planet detections in RV-blind searches through optimizing the scheduling of  observations. It is an open source code written in Python language available to the community. Bayesian adaptative scheduling algorithms have been already demonstrated to be powerful in previous works (e.g., \citealt{2004AIPC..707..330L}; \citealt{2008AJ....135.1008F}; \citealt{2011arXiv1108.0020L}). They propose improving observational efficiency on-the-fly by developing algorithms composed of two steps: inference and decision. The former quantifies the knowledge acquired with the available data and the latter chooses the optimum date depending on the scientific goal (more information in Sect.\,\ref{sec:beta}) and based on the predictions. Particularly, \citealt{2008AJ....135.1008F} shows through simulations that their algorithm has the potential not only to increase planet detections, but also to increase the sensitivity compared to conventional strategies. 
  
 \texttt{KOBEsim} was developed to enhance the planet detection in the K-dwarfs Orbited By habitable
Exoplanets (KOBE) experiment,\footnote{\url{https://kobe.caha.es/}}  a blind-search RV survey devoted to the hunt for rocky and potentially habitable exoplanets around K-dwarf stars (\citealt{2022A&A...667A.102L}). KOBE is a legacy program of the Calar Alto Observatory (CAHA; Almería, Spain), making use of the Calar Alto high-Resolution search
for M dwarfs with Exoearths with Near-infrared and optical Echelle Spectrographs (CARMENES), a fiber-fed échelle spectrograph (\citealt{2020SPIE11447E..3CQ})
  at the 3.5 m telescope. KOBE observations began in  January 2021 and will be monitoring 50 late K-dwarf stars over five consecutive semesters. The main goal of this experiment is to bridge the gap between G  and M dwarfs  in the search for planets within the habitable zone (HZ), a parameter space that has been barely explored (see Fig. 1 in \citealt{2022A&A...667A.102L}).

 This paper is organized as follows. In Sect. \ref{sec:meth} we describe the methodology, and  provide a description of the architecture of the \texttt{KOBEsim} code. In Sect. \ref{sec:test} we show the results of testing the strategy by applying the algorithm to simulated and real datasets. We also briefly illustrate how the data simulation is performed, we show the efficiency of \texttt{KOBEsim} in detecting planets of different masses by comparison with monotonic cadence strategies, and we study the case of targeting false positive periodicities. Finally, in Sect. \ref{sec:concl} we summarize the results and present our conclusions.

\section{Methodology}
\label{sec:meth}

 The proposed observational strategy consists of two steps (see Fig. \ref{fig:diagram}). First, the star is monitored with the usual survey strategy; for example, in the KOBE experiment the targets are observed with a cadence of $\sim$10\,\% with respect to the orbital period that a planet would have in the middle of the HZ. This step lasts until there are enough RV data gathered ($n$) to see an emerging peak in the periodogram. Second, that period ($P_{\rm peak}$) is pursued by the \texttt{KOBEsim} algorithm. By ranking all the possible observing dates, the algorithm proposes the optimum next observing night. In Sect. \ref{sec:test} we show that this turns out to be faster than a monotonic cadence strategy (i.e., continuing to use the strategy followed in the first step) to determine whether $P_{\rm peak}$ is due to a strictly periodic signal (induced by the presence of a planet).

 \subsection{Statistical framework}

In this work we assume the concept of planet detection based on a Bayes factor threshold for two competing models: the null hypothesis $H_0$ where the parameters ($\theta$) do not include any Keplerian orbit, against the alternative $H_1$, where the parameters include a Keplerian orbit. In statistical notation we write 
\begin{equation}
    H_0: \theta \in \Theta_0 \quad \mathrm{vs.} \quad H_1: \theta \in \Theta_1,
\end{equation}
where $\Theta_i$ represents a restricted range of values where the model includes a planet or not. In the first step of \texttt{KOBEsim}, we infer the parameters that describe these competing models (detailed in Sect. \ref{sec:estimating}). For this purpose we model the $j$-th RV observation as
\begin{equation}\label{eq:vj}
    \mathrm{v}_j (t_j) = V_{\mathrm{sys}} + \mathcal{K}(t_j) + E_j,
\end{equation}
where $V_{\rm sys}$ is the systemic RV, $\mathcal{K}(t_j)$ is the equation of RV corresponding to a Keplerian signal depending on the time of the measurement $t_j$, and $E_j$ is the noise contribution obtained from a Gaussian distribution with zero mean and variance $\sigma_j^2 + S_j^2$. The value $\sigma_j^2$ is the uncertainty associated with the $j$-th measurement due to the photon noise, and $S_j^2$ are variances of unmodeled sources of noise. In the particular case of a RV signal induced by one planet ($H_1$), $\mathcal{K}(t_j)$ can be written as
\begin{equation}\label{eq:keplerian}
    \mathcal{K}(t_j) = K \left[ \cos{ \left( \nu( t_j,\, e,\, P,\, t_0) + \omega \right)} + e \cos{\omega} \right],
\end{equation}
where $K$ is the RV semi-amplitude, $\nu$ is the true anomaly of the planet (angle between the periastron and the position of the planet in the elliptical orbit measured from the central star), \textit{e} is its eccentricity (degree of deviation of the elliptical orbit from a circle), $P$ is the orbital period of the planet, $t_0$ is the inferior conjunction time, and $\omega$ is the argument of the periastron (angular distance between the line of nodes and the periastron). Thus, $\mathrm{v_j}$ depends on the following parameters:\begin{equation}\label{eq:theta1}
\Theta_1 = (V_{\mathrm{sys}},\, K,\,P,\, t_0,\, e,\, \omega,\, S_j)
.\end{equation}
Meanwhile, $\mathcal{K}(t_j)$ is zero for the null hypothesis $H_0$, thus the parameters are
\begin{equation}
    \Theta_0 = (V_{\mathrm{sys}},\, S_j)
.\end{equation}
We estimate the posterior probability of each hypothesis using the Bayes relation
\begin{equation}
     P(H_i | {\bf D}) = \frac{P(H_i) \mathcal{Z}_i}{P(H_i) \mathcal{Z}_i + P(H_j) \mathcal{Z}_j},
\end{equation}
where $\mathcal{Z}_i$ is the evidence (i.e., marginal likelihood) under the hypothesis $i$, and \textbf{D} is an array of datapoints. Thus, the ratio for the competing models (the posterior odds) is
\begin{equation}
    \frac{P(H_1 | {\bf D})}{P(H_0 | {\bf D})} = \frac{P(H_1)}{P(H_0) } B_{10},
\end{equation}
where $B_{10}$ is the evidence (or marginal likelihood) ratio,  called the Bayes factor. Assuming the same prior probability of a star hosting one or no planets (i.e., $P(H_1)/P(H_0)\,=\,1$), we can use $B_{10}$ as a metric to analyze how significant one model is compared to another for a given dataset. An intrinsic feature of this mathematical construction is  Occam's razor (e.g., \citealt{2003itil.book.....M}; \citealt{2019PASA...36...10T}), which  penalizes the most complex models (i.e., those with more parameters). In this work we set the limit to consider a planet detection at $\ln{\left(B_{10}\right)} > 6$, which is a conservative criterion since it is four times greater than the evidence \citet{1961Ox...J} proposed as decisive.

\subsection{Estimating the evidence of the models}
\label{sec:estimating}

\setlength{\tabcolsep}{12pt}
\begin{table}[h]
    \centering{
      \caption[]{Prior distributions to perform the MCMC fit.}
    \label{tab:prior}
    \begin{tabular}{@{}ccc@{}}
    \hline \hline
    Parameter & Prior& Units \\ \hline
    $V_{\rm sys}$             & $\mathcal{U}\left(-10^{5}, 10^{5}\right)$      & m\,s$^{-1}$                                 \\
    ${K}$     & $\mathcal{U}\left(0, 10^{4}\right)$   & m\,s$^{-1}$                                       \\
    $P$  & $\mathcal{G}$($P_{\rm peak}$, 4)             & d                                \\
    $t_0$      & $\mathcal{G}$(t$_{0,\rm input}$, 4) or $\mathcal{U}\left(t_{1}, t_{1} + P_{\rm peak}\right)$  & d\\
    $\sqrt{e}\cos{\omega}$      & $\mathcal{G}_t\left(0, 0.3\right)$  & -\\
    $\sqrt{e}\sin{\omega}$      & $\mathcal{G}_t\left(0, 0.3\right)$  & -\\
    $S_j$    & $\mathcal{U}\left(0, 10^{2}\right)$      & m\,s$^{-1}$ \\ \hline
    \end{tabular}}
   \end{table}

Once we have a predominant signal ($P_{\rm peak}$) we are ready to run \texttt{KOBEsim}. First, it derives the set of parameters from the data  for both the null ($H_0$) and planet ($H_1$) hypotheses (corresponding to Fig. \ref{fig:diagram} panel \textit{a}). We explore the parameter space and sample the posterior distribution by using the Markov chain Monte Carlo (MCMC) affine invariant ensemble sampler \texttt{emcee} (\citealt{2013ascl.soft03002F}). To compute the Keplerian in the H$_1$ model, Eq. (\ref{eq:keplerian}), we use the python module \texttt{RadVel} (\citealt{2018PASP..130d4504F}). Considering normally distributed data around the theoretical value of the model, our selection of likelihood function is a Gaussian-noise
\begingroup
\fontsize{9.8pt}{10pt}
\begin{equation}
    -2 \ln \mathcal{L}({\bf D} | \theta) = Q + \sum_{j=1}^n \ln \left( \sigma_j^2 + S_j^2 \right) + \sum_{j=1}^n \frac{\left[ V_{\mathrm{sys}} + \mathcal{K} (t_j) - \mathrm{v}_j \right] ^2}{\sigma_j^2 + S_j^2},
\end{equation}
\endgroup
where $\bf D$ is our RV measurement and its associated uncertainties (${\bf D}\,=\,\{(\mathrm{v_j}, \sigma_j)\}_{j=1}^n$), and $Q$ is a constant.
We assume the prior distributions to be uniform, except for those parameters we are more informed about, for which we select a narrow normal distribution (e.g., $P_{\rm peak}$). The prior distributions used for each parameter are shown in Table \ref{tab:prior}, where $t_{0,\rm input}$ is the value of $t_0$ that can be optionally given as input, $t_1$ is the first day that the target was observed, and $\mathcal{G}_t$ corresponds to a truncated Gaussian between -1 and 1. To sample the parameter space, we employ four times the number of parameters of the model and $2 \times 10^{4}$ steps in each of them\footnote{Both the number of walkers and steps can be customized by the user; see Table \ref{tab:inputs}.}. To speed up the convergence, we start new chains in a ball around the best solutions from the previous sampling with half the  number of steps. Following the criterion suggested in the documentation of \texttt{emcee}\footnote{\url{https://emcee.readthedocs.io/en/stable/tutorials/autocorr/}}, we consider our sampling to be  successful when the chains are longer than 50 times the autocorrelation time.

To calculate the Bayes factor metric we employ the \texttt{bayev} code (\citealt{2016A&A...585A.134D}), which uses the estimator defined in \citet{PERRAKIS201454}. Giving as input a representative fraction of the marginalized posterior distributions provided by \texttt{emcee}, the likelihood function, and the priors, we obtain the $\ln(\mathcal{Z})$ distribution. The authors of the \texttt{bayev} code estimate the uncertainty of this distribution repeatedly reshuffling the joint posterior sample to produce new samples. In our case we opt for  just one fraction of the distributions for computational time reasons since we  checked that the chosen fraction (15\,\% of the iterations) is representative enough and the standard deviation of the resulting distributions does not change significantly (around 20\,\%). Carrying out this procedure for the two competing models, and considering a priori the same probability for  each hypothesis, $P(H_1)\,=\,P(H_0)$, we obtain an estimation of $\ln(B_{10})$ with its associated uncertainty.\\
\subsection{Forecasting the optimum observing date}
\label{sec:beta}

In the second step of the code we use the existing data to select the future date that most (or more optimally) increases the evidence of the planet model at the targeted periodicity. For this  purpose, we predict and compare the expected increase in the $\ln(B_{10})$ for each candidate date in our schedule. The Bayes factor is our metric for  measuring the gain, which in decision theory is called the utility function. This quantity is commonly used in RV datasets to test different hypotheses, and thus can be used to claim a planet candidate  (e.g., \citealt{2020A&A...642A.121L}; \citealt{2020MNRAS.499.5004M}; \citealt{2022A&A...658A.115F}). Previous works on Bayesian adaptative schedulers adopt other utility functions based on their goals. For instance, \citealt{2011arXiv1108.0020L} opt for the Shannon entropy as they aim to improve the efficiency in constraining the parameter inference, and this utility function is used to  reduce the posterior uncertainties.

The prediction is the stage illustrated in Fig. \ref{fig:diagram} panel \textit{b}. To find the candidate dates we divide the period into a total of $N_{\rm phases}$ orbital sub-phases (i.e., spliting the orbital phase $N_{\rm phases}$ times). We choose the next assigned date from the schedule at the telescope that matches each sub-phase. In this selection of the candidate dates, we take into account twilights, altitude of the target, and exposure time (see Sect. \ref{sec:sim} for further details). Next, using the whole posterior probability distributions inferred for our model parameters through the MCMC algorithm (obtained in the stage shown in Fig. \ref{fig:diagram} panel \textit{a}),  we sample the posterior predictive distribution for the RVs at each potential observing date by using Eq. (\ref{eq:vj}). We take the uncertainty of the predicted RV from the quadratic sum of the standard deviation of the predictive distribution (uncertainty due to the parameter inference) and a random value from a normal distribution of same mean and standard deviation as the uncertainities ($\sigma_j$) of the \textit{n} RV datapoints already gathered (simulating the expected photon noise).

Running again \texttt{emcee} and \texttt{bayev} over each of the datasets (including one additional datapoint corresponding to each predicted RV at a proposed date), we end up with an estimation of $\Delta\ln(B_{10})$ for each of the proposed future dates. \texttt{KOBEsim} sorts the tested dates according to the utility function, giving the maximum priority to the highest $\Delta\ln(B_{10})$ (stage illustrated in Fig. \ref{fig:diagram} panel \textit{c}).

For long targeted periodicities, the largest $\ln(B_{10})$ increase may occur at a very distant date, which is against the efficiency of the observations. Consequently, the detection using \texttt{KOBEsim} could require a long time span despite needing a lower number of measurements. To prevent this situation, we introduce a weight to the utility function with the shape of a density function of a beta distribution, such that
    \begin{equation}\label{eq:beta}
        \Delta \ln{(B_{10})} = \beta(\Delta t, a, b)\left[ \ln{(B_{10, n+1})} - \ln{(B_{10, n})}\right],
    \end{equation}
where $\Delta t$ is the difference in days between the new proposed observation and the current date (normalized on all candidate dates), and $n$ denotes the number of gathered datapoints at the moment of running \texttt{KOBEsim}. We choose the arguments of the beta distribution in such a way that in order to end up with a time gap (i.e., time between the last observation and the next one)
greater than 40\,\% of the targeted periodicity, the increase in the $\ln(B_{10})$ should be at least five times greater than for the next date in the priority list (\textit{a}\,=\,1 and \textit{b}\,=\,5). The beta distribution weight is an optional parameter of \texttt{KOBEsim} and it is activated by default.

\section{Results}
\label{sec:test}

 In this section, we show different analyses to test the algorithm against simulated (Sect. \ref{sec:sim}) and real  (Sect. \ref{sec:real}) datasets.

\subsection{Testing \texttt{KOBEsim} on simulated datasets}
\label{sec:sim}

\subsubsection{Simulated data}
\label{sec:simdata}

To generate synthetic data we first had to simulate the observing dates. The observing time must be between the astronomical twilights from the observatory location (i.e., when the Sun is at 18 degrees below the horizon) and the elevation of the target over the horizon must be greater than a given minimum altitude during the exposure time to avoid large chromatic distortions and extinction due to the atmosphere (e.g., \citealt{2011A&A...525A.140D}). Through the following examples, we consider a probability of 55\,\% for a night to be assigned to the project, and we assume that 70\,\% of the nights meet the appropriate weather conditions to perform the observations.

Second, we simulated the RV measurements. The $V_{\rm sys}$, $P$, and $t_0$ were drawn from uniform distributions.  For this example, we decided the boundary values for $P$ to cover the properties of the KOBE sample (orbital periods inside the HZ of late-type K dwarfs calculated as defined by \citealt{2014ApJ...787L..29K}). We calculated the RV semi-amplitude as
\begin{equation} \label{eq:k}
    K = \frac{1}{\sqrt{1 - e^2}} \frac{m_p \; \sin{ \; i}}{(m_p + M_{\star})^{2/3}} \left( \frac{2 \pi G}{P} \right)^{1/3},
\end{equation}
which requires four more parameters: the planetary mass ($m_{p}$), the orbital inclination ($i$), the eccentricity ($e$), and the stellar mass ($M_{\star}$).  We note that $G$ represents the gravitational constant. We assume\,$M_{\star}$ to be equal to 0.5\,$M_{\odot}$ since it is the mean value of the KOBE sample (spectral types from K5 to M0). For $m_{p}$ we explore different scientific cases (5, 10, 20, and 60\,$M_{\oplus}$). Finally, to simplify the problem, we assume an edge-on system ($i$\,=\,90$^{\circ}$) and a circular orbit ($e\,=\,0$). In Table \ref{tab:rvsim} we show the parameter selection used for this test. In this case $V_{\rm sys}$, $P$, and $t_0$ has a value within the indicated range, while the  $m_p$ value is selected from the four given options. We add white noise to the mock RVs through a normal distribution with a mean of 0 m\,s$^{-1}$ and a standard deviation of 3 m\,s$^{-1}$ mimicking the instrumental noise (conservative values for the CARMENES instrument). We consider a Gaussian uncertainty associated with the simulated RV data, with a mean of 3 m\,s$^{-1}$ and a standard deviation of 0.3 m\,s$^{-1}$. Since the aim of  \texttt{KOBEsim} is to  improve the efficiency of Keplerian modulation detection, we work under the assumption of a well-characterized stellar activity (e.g., \citealt{2011A&A...525A.140D}; \citealt{2017A&A...606A.107O}). Therefore, we do not include red noise in our simulated data, and it is a caveat the user must bear in mind.

\setlength{\tabcolsep}{8pt}
\begin{table}
\caption[]{Values of the orbital parameters used for the RV simulation.}
\label{tab:rvsim}
\centering
\begin{tabular}{c c c}
\hline\hline
 Parameter & Value for the simulation & Units\\ \hline
$V_{\rm sys}$ & $\mathcal{U}\left( -10, 10 \right)$& m\,s$^{-1}$ \\
$P$ &  $\mathcal{U}\left( 50, 120\right)$ & d\\
$t_{0}$ &  $\mathcal{U}\left( t_1, t_1 + P\right)$& d\\
$M_{\star}$ &  0.5 &$M_{\odot}$\\
$m_p$  &   (5, 10, 20, 60)&\,$M_{\oplus}$ \\
$i$ & 90 & deg\\
$e$ & 0 & -\\
\hline
\end{tabular}
\end{table}

\begin{figure*}
  \begin{center}
      \subfigure{\includegraphics[width=80mm]{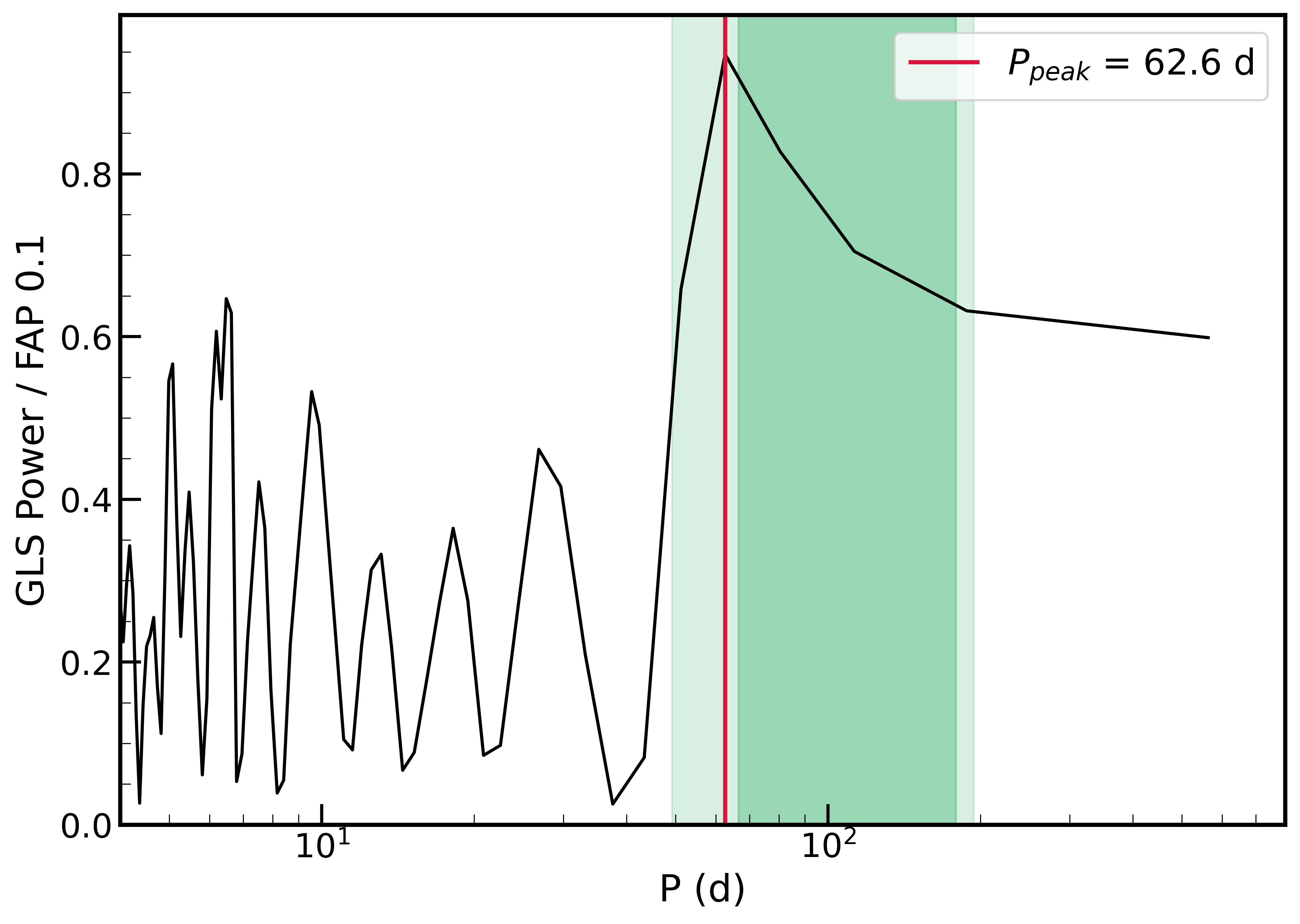}}\hspace{0.2cm}
      \subfigure{\includegraphics[width=92mm]{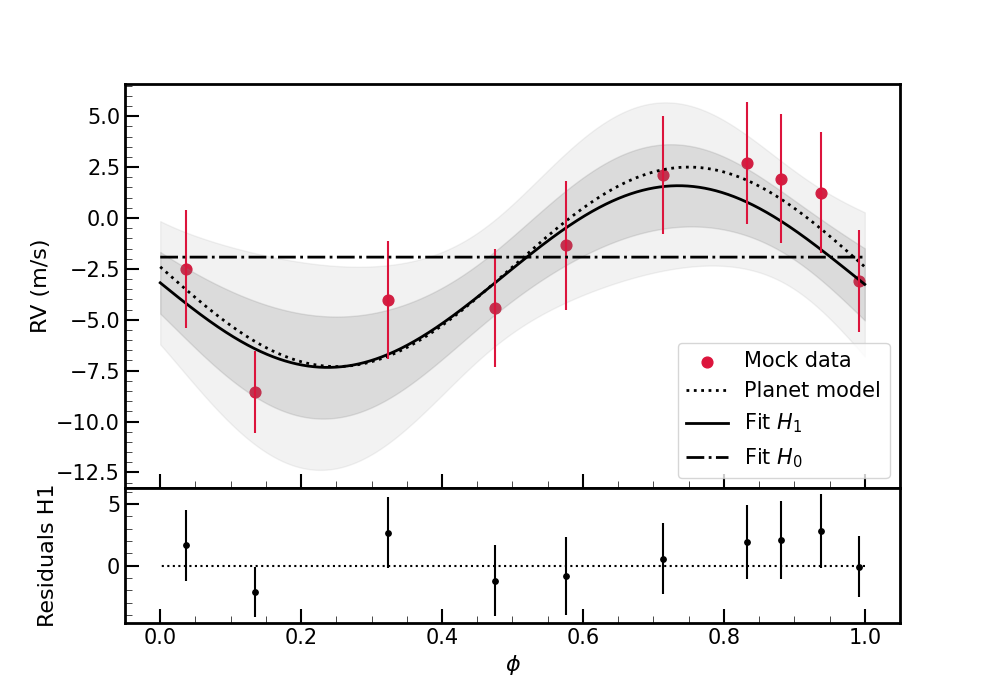}}
  \end{center}
 \caption{Simulation of ten RV datapoints to test \texttt{KOBEsim}. \textit{Left}: GLS periodogram showing an emerging period at $P_{\rm peak}$\,=\,62.6 d nearly reaching a false alarm probability (FAP) of 0.1 (\textit{y}-axis equal to 1). The lighter green region corresponds to the optimistic HZ, the darker region to the conservative HZ as defined in \citet{2014ApJ...787L..29K}. \textit{Right}:  Fit of the simulated RV data. The solid black line is the fit for one-planet hypothesis ($H_1$); the dash-dotted  line is  fit for the null hypothesis ($H_0$); and  the dotted line is the  true model used to simulate the datapoints (in red). The shaded region shows the confidence interval at 1$\sigma$ (dark gray) and 2$\sigma$ (light gray) for the hypothesis with a planet. The lower panel shows the residuals for the one-planet hypothesis fit.}
 \label{fig:fit_simul}
\end{figure*}

\begin{figure*}
  \begin{center}
     \subfigure{\includegraphics[width=130mm]{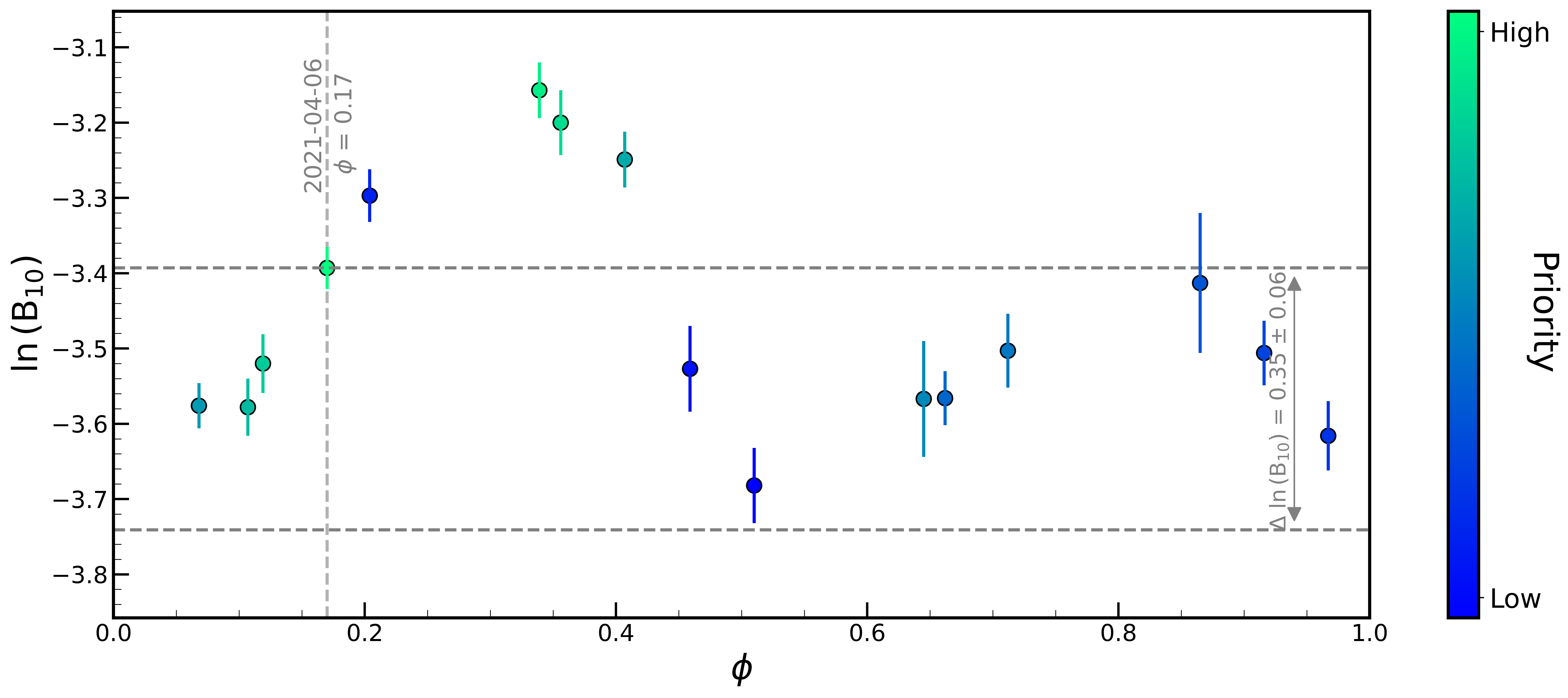}}
  \end{center}
     \caption{\texttt{KOBEsim} output figure. The $y$-axis shows the logarithm of the expected Bayes factor and the $x$-axis shows the orbital phase. The vertical dashed line gives the phase selected as the optimal option, and gives its corresponding date. The horizontal dashed lines indicate the initial $\ln(B_{10})$ (lower line) and the expected value at the selected phase (upper line). The increment is shown with a gray arrow. The color-coding shows the priority order for selecting the next observing date.}
  \label{fig:KOBEsim_simul}
\end{figure*}

\subsubsection{Running \texttt{KOBEsim}}
\label{sec:simuldaily}
\noindent \texttt{KOBEsim} should always be executed after obtaining a new datapoint for a given target. In this way the state of knowledge is updated and provides a list of the next possible observing dates ordered by priority as detailed in Sect. \ref{sec:meth}. We use the above-described simulation tool to generate simulated observations taking into account the visibility from CAHA for one of the KOBE targets. We generated the first ten observations assuming a model in which $m_p$\,=\,20\,$M_{\oplus}$ and $P$\,=\,59\,d. In Fig. \ref{fig:fit_simul} we show the GLS periodogram (left panel) and the phase-folded RV curve after the parameter inference is performed for both competing models (right panel). Thus, $P_{\rm peak}$\,=\,62.6 d is our target periodicity to test with \texttt{KOBEsim} (compatible with the true period $P$ used in the model as the prior is Gaussian with a $\sigma$\,=\,4 d; see Table \ref{tab:prior}). The goal now is to predict the best observing date for the next datapoint (the 11th in the time series) in order to speed up a possible planet detection at that periodicity.

The two models we employ to fit the data allow us to quantify how well supported the planet model is against the null hypothesis with the available data. In this case we obtain $\ln{(B_{10,\rm initial})}$\,=\,--3.74\,$\pm$\,0.05. \texttt{KOBEsim} now generates a new synthetic datapoint based on the inferred planetary orbital and physical parameters and adds it to the previous datapoints, calculating the increase in $\ln(B_{10})$. This process is repeated independently for different orbital phases. Here we compare a total of 20 candidate orbital phases ($N_{\rm phases}$\,=\,20), which for this simulated planet is equivalent to comparing dates around three days apart from each other. In this particular simulation, we imposed the constraint that if the next matching date is more than three months away from the current date, that sub-phase is discarded\footnote{The \textit{maximum days apart} parameter can be set by the user; see Table \ref{tab:inputs}.}: since the KOBE program monitors 50 targets and it was awarded with $\sim$55\,\% of the nights, the chances of observing the target are high, and waiting that long would imply a waste of telescope time. Once all of the candidate dates are calculated, \texttt{KOBEsim} computes which  is the best given the current data and assuming the input periodicity. The execution of the code takes $\sim$32 s/phase and is repeated $N_{\rm phases}$ + 1 times, which leads to a total runtime of $\sim$15 min.

In Fig. \ref{fig:KOBEsim_simul} we show the output plot returned by the algorithm. The difference in $\ln(B_{10})$ of the one-planet model against the null hypothesis is shown for each orbital phase. The figure shows that the algorithm generated fewer than 20 candidate points. This is caused by the impossibility of observing the target over the next three months, either because the corresponding dates are not assigned to the program or because the star has already set. As in this example, we   use the beta distribution, meaning that  the criterion for selecting the best observing date is not the one that maximizes the increase in $\ln(B_{10})$. Instead, \texttt{KOBEsim} maximizes the weighted $\Delta\ln(B_{10})$ given by Eq. (\ref{eq:beta}), thus finding a trade-off between efficiency and the time we have to wait until the next observation. In this example, we find out that this occurs for an orbital phase around $\phi$\,=\,0.17, reaching $\Delta\ln{(B_{10})}$\,=\,+0.35\,$\pm$\,0.06. In Appendix \ref{sec:A} we show the corresponding csv output (Table \ref{tab:csv1}). In view of Fig. \ref{fig:KOBEsim_simul}, every candidate observation will lead to a positive increase in $\ln(B_{10})$ or it will be maintained at the initial value. This may not occur if we are targeting the wrong period, as we discuss in Sect. \ref{sec:false}.

\subsubsection{Efficiency}
\label{sec:eff}

To study the efficiency of \texttt{KOBEsim}, we estimate how long it would take to detect planets of different masses within the HZ for the particular case of the KOBE experiment. We compare this with the time and the number of observations that a monotonic cadence approach (i.e., obtaining observations every $N$ days) would require. For this purpose, we simulate the future observations for a given target, cumulatively, until we obtain enough evidence from the one-planet model over the null hypothesis to claim the detection. We  set this limit at $\ln{(B_{10})} > 6$. We performed this procedure for both \texttt{KOBEsim} modes (with and without the beta distribution), and spacing the observations with the cadence assigned to our testing target (6\,$\pm\,2$\,d) as long as that day meets good weather conditions and if it is a date granted for the project (see Sect. \ref{sec:simdata}), otherwise it is postponed to the next plausible day. Hereafter, we refer to the  strategies as \texttt{K} (for \texttt{KOBEsim}), \texttt{K$_{\beta}$} (\texttt{KOBEsim} beta), and MC (monotonic cadence). We note that the simulations of new observations (corresponding to the right circle in Fig. \ref{fig:diagram}) are computed independently from \texttt{KOBEsim} and with the sole idea of testing the efficiency of our methodology. Thus, after deciding the optimum next date, we predicted again the RV value at the corresponding orbital phase. Contrary to the  \texttt{KOBEsim} simulation stage (Fig. \ref{fig:diagram} panel \textit{b}),  in this case we included  white noise,  as explained in Sect. \ref{sec:simdata}, but we did not include in the associated uncertainty the component due to the inference (the standard deviation from the RV predictive distribution) since the sources of uncertainty in real observations are only the jitter ($E_j$) and the photon noise ($\sigma_j$).

\begin{figure*}
  \begin{center}
     \subfigure{\includegraphics[width=150mm]{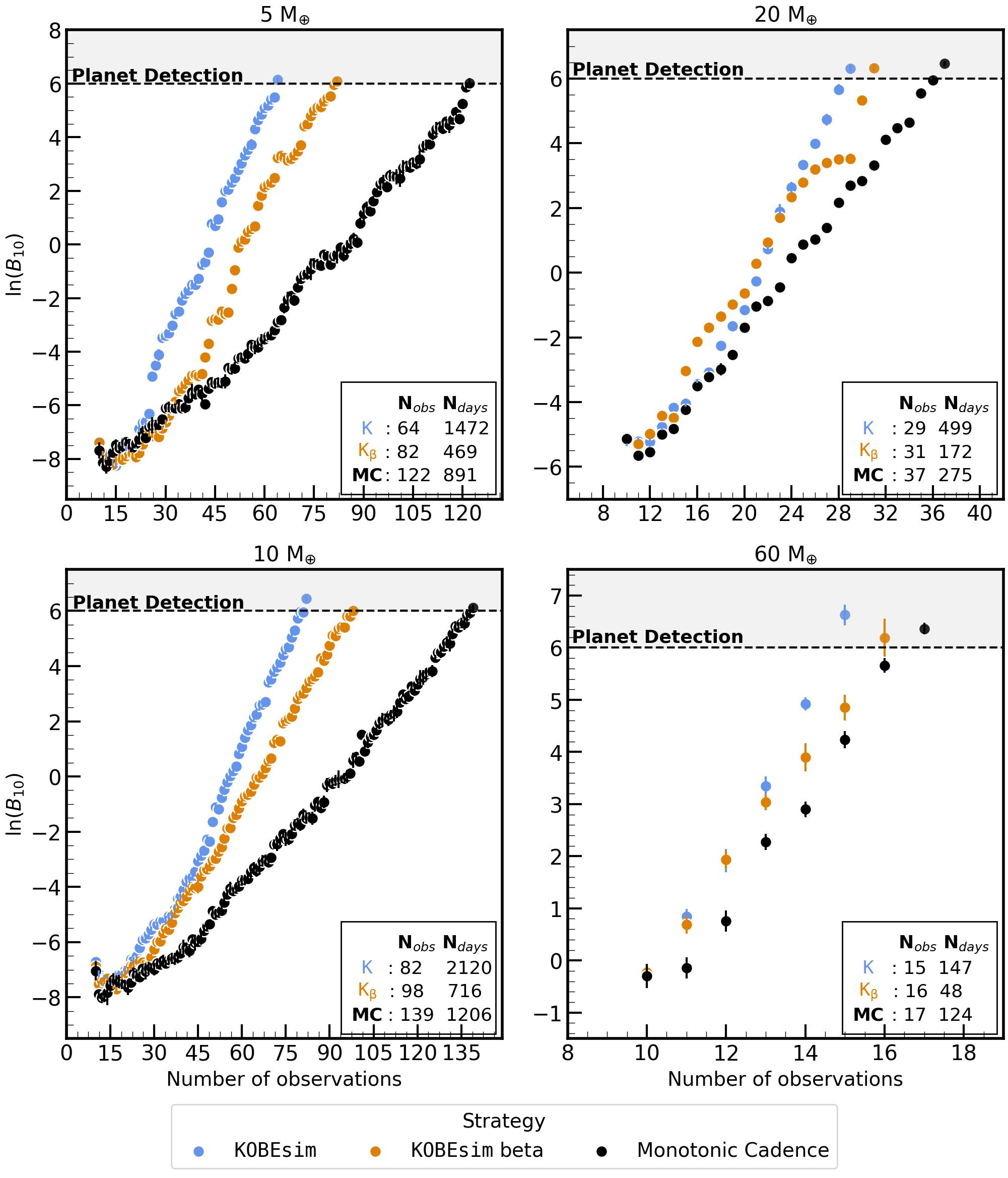}}
 \end{center}
\caption{Prediction in the evolution of the logarithm of the Bayes factor for simulated planets of 5, 10, 20, and 60\,$M_{\oplus}$ at $P$\,=\,59 d. The period targeted with \texttt{KOBEsim} is $P_{\rm peak}$\,=\,57.54 d. The number of observations and the time it would take to detect the planet are compared using three different strategies: \texttt{KOBEsim} (K), \texttt{KOBEsim} beta (K$_{\beta}$), and spacing the observations at a fixed cadence (MC) of 6 d.}
\label{fig:seqall}
\end{figure*}

In Fig. \ref{fig:seqall} we show the results for planets of 5, 10, 20, and 60\,$M_{\oplus}$ using $P$\,=\,59 d and $P_{\rm peak}$\,=\,62.6\,d. In this section we do not update $P_{\rm peak}$; we keep it fixed to run \texttt{KOBEsim} at each iteration. For all of these planetary masses except for 5\,$M_{\oplus}$, we considered the visibility of our testing target. As the precision of CARMENES in its optical arm is around 1 m\,s$^{-1}$, a planet of 5\,$M_{\oplus}$ corresponds to the detectability limit case for the KOBE experiment (RV semi-amplitude within the HZ of late K dwarfs between 0.72 and 1.15 m\,s$^{-1}$). Since such a detection is very demanding, we simulated these measurements considering a circumpolar star, and thus every night of the year can be used to collect data if the weather conditions are favorable. In addition, to achieve a detection of this kind in the time that the KOBE program lasts, it would be necessary to increase the exposure time (thereby increasing the signal-to-noise ratio to reduce the uncertainty). In practice, this would only be feasible with the brightest targets as the maximum exposure time allowed by the CARMENES instrument is 1800\,s. For this reason for 10, 20, and 60\,$M_{\oplus}$ (see Fig. \ref{fig:seqall}), we considered a conservative uncertainty following a normal distribution with a mean of 3 m\,s$^{-1}$, whereas for 5\,$M_{\oplus}$ we reduced the mean to 1.5 m\,s$^{-1}$. We started with ten initial datapoints. In every planetary mass case, we see the gradual increase in $\ln{(B_{10})}$ that each strategy follows as new observations are added. We   checked that different initial sets of simulated data (i.e., varying the white noise in the RV, the first observing date, and phase coverage, but keeping the same time span and number of measurements) did not significantly change the results. Particularly, in Fig. \ref{fig:seqall} the standard deviation in the y-axis ($\ln{(B_{10})}$) is below 1 for every number of observations and the relative error in the corresponding slopes are 6\,\%. At  first glance,   it is clear that the number of   observations needed to confirm the planetary signal is greatly reduced when using the \texttt{KOBEsim} approach, especially for the less massive planets. Furthermore, the number of days invested in these observations is also greatly reduced even for the most massive planets when using \texttt{K$_{\beta}$}.

\setlength{\tabcolsep}{3pt}
\begin{table}
\caption[]{Improvement achieved using \texttt{KOBEsim} beta ($\tt{K_{\beta}}$) in comparison with a monotonic cadence (MC) strategy for simulated datasets.}
\label{tab:improve}
\centering
\begin{tabular}{c c c}

&  {\tt KOBEsim} beta Improvement \\
\hline\hline
$m_p$ $(M_{\oplus})$ & Datapoints& Time span\\
\hline
5  & 33\,\%                                 & 47\,\%                                 \\
10                                   & 29\,\%                                 & 41\,\%                                 \\
20                                & 16\,\%                                 & 38\,\%                             \\
60                                   & 0\,\%                                  & 61\,\%                             \\
\hline
\end{tabular}
\end{table}

The large time span required for detecting the two most massive simulated planets when applying the  \texttt{K} strategy (147\,d for the 60\,$M_{\oplus}$ simulation) in comparison with both \texttt{K$_{\beta}$} (48\,d) and MC (124\,d) strategies, can be explained by the altitude of the testing star. This target sets over three months after starting the simulations. Therefore, the detection in the former strategy, unlike the other two, is  postponed until after the target rises again. This example highlights the importance of using the beta distribution since the efficiency of the observations requires finding a compromise between time span and number of measurements. These four simulations show that \texttt{K$_{\beta}$} greatly reduces the time span, and the gain in terms of number of observations is nearly as good as in \texttt{K}.

The efficiency gain by means of \texttt{K$_{\beta}$} in comparison with MC strategy is shown in Table \ref{tab:improve}. It collects the key information to support the strength of this algorithm in the context of blind-search surveys. In terms of the number of observations, the improvement varies from 16\,$\%$ for the heavy planets (20\,$M_{\oplus}$) to 29\,--\,33\,$\%$ for the light planets (5\,--\,10\,$M_{\oplus}$). Furthermore, regarding the improvement in the number of days, \texttt{KOBEsim} can be  decisive even  for the  high-mass planets mentioned above since in this particular case we achieved an improvement of 38\,--\,61\,$\%$ (20\,--\,61\,$M_{\oplus}$). Finally, the most impressive improvement is related to the number of days for low-mass planets, reducing the time span by 41\,--\,47\,$\%$ (5\,--\,10\,$M_{\oplus}$). This could increase the speed of detection of rocky planets within the HZ of the parent star by nearly a factor of 2, provided the RV modulations due to stellar activity are well known.

\subsubsection{False detections}
\label{sec:false}

  We test the behavior of \texttt{KOBEsim} when pursuing spurious periodicities caused by signals either behaving stochastically or not induced by Keplerian sources. If we have very few datapoints at the time of period selection, it may occur that the period pursued does not correspond to a planet signal. It is also possible that we select a periodicity resulting from stellar activity mimicking the wobble of the star when it has an orbiting planet (e.g., \citealt{2001A&A...379..279Q}; \citealt{2010EAS....42..131F}; \citealt{2014A&A...566A..35S}). These possible scenarios raise the question of how \texttt{KOBEsim} behaves against a spurious period.

 \begin{figure*}
  \begin{center}
     \subfigure{\includegraphics[width=160mm]{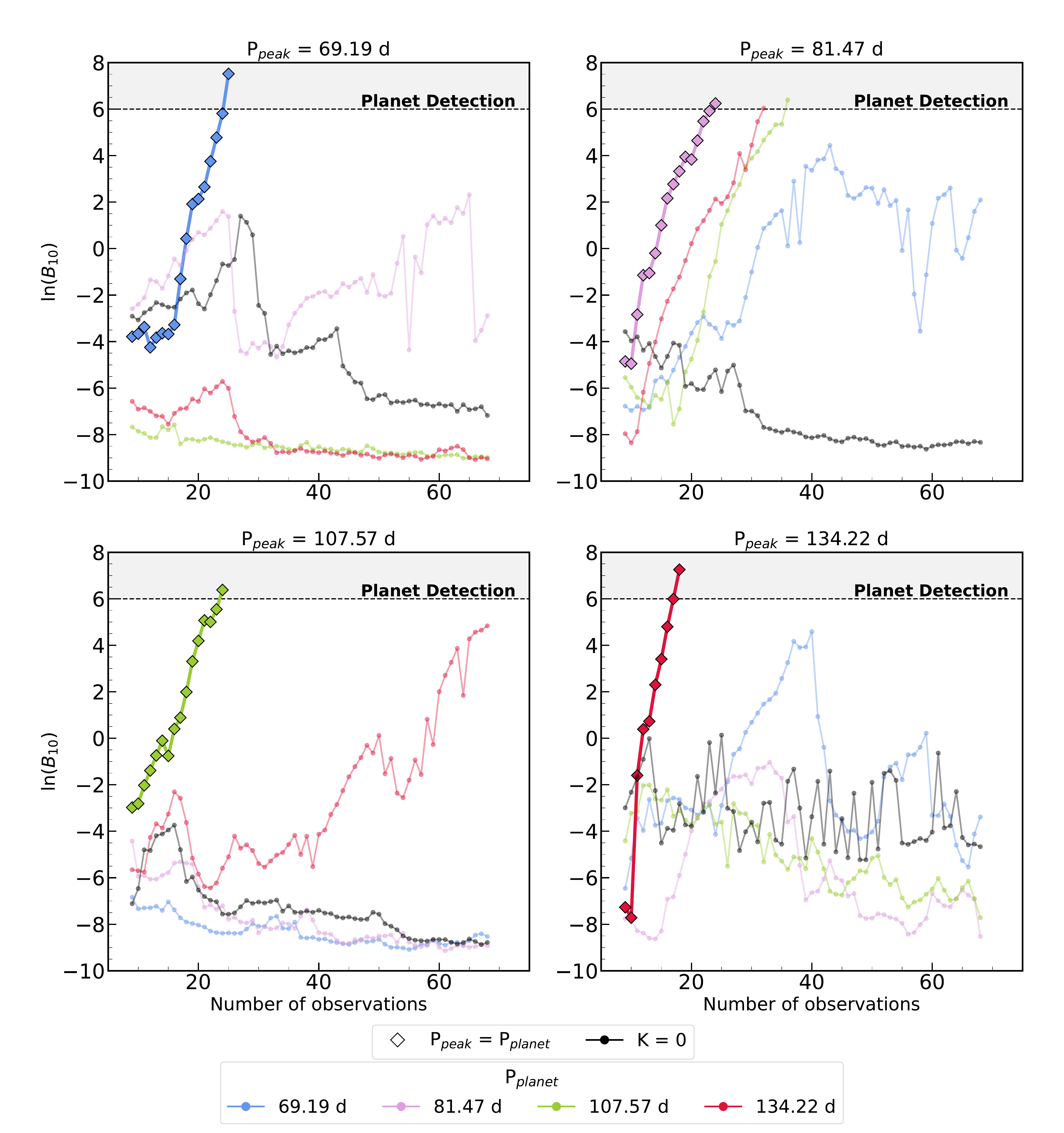}}
  \end{center}
 \caption{Prediction in the evolution of the logarithm of the Bayes factor when targeting with \texttt{KOBEsim} a period $P_{\rm peak}$ (each chart), but the signal is induced by a planet at $P_{\rm planet}$ (see legend for colors). All the cases induce a signal of semi-amplitude $K$\,=\,6\,m\,s$^{-1}$. The solid lines are the cases in which \texttt{KOBEsim} is pursuing the correct period ($P_{\rm peak}$\,=\,$P_{\rm planet}$). The black line corresponds to the active-star case. After the first ten observations it is assumed that the signal is turned off ($K$\,=\,0).}
 \label{fig:Psp}
\end{figure*}

  \begin{figure}
  \begin{center}
     \subfigure{\includegraphics[width=75mm]{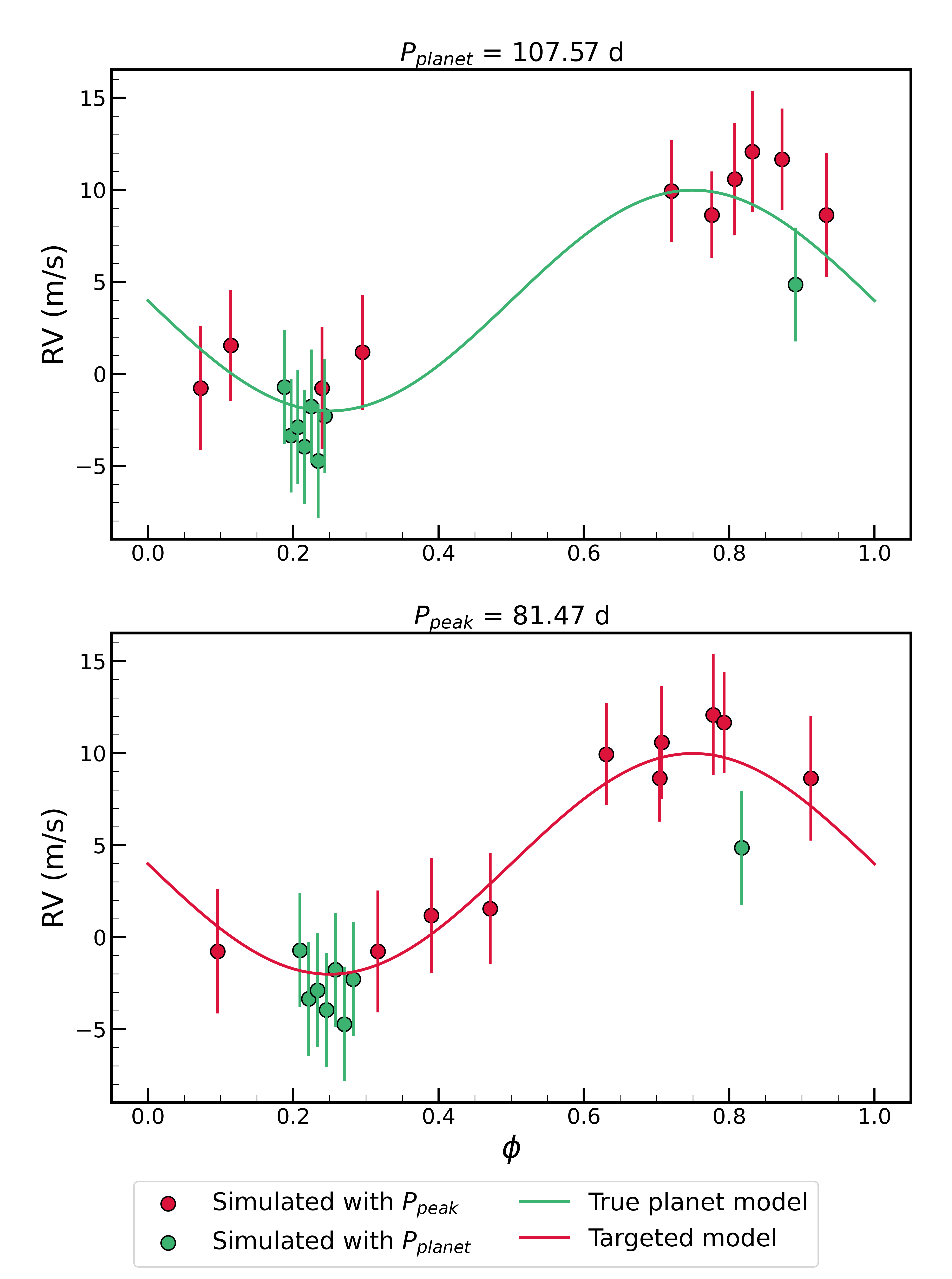}}
  \end{center}
 \caption{Phase-folded RV curve of simulated data. \textit{Top}: Assuming the true orbital period ($P_{\rm planet}$). In red are indicated the first observations, generated with $P_{\rm peak}$ constrained to be compatible with the true signal within 1$\sigma$. \textit{Bottom}: Assuming the incorrect period, the one used to perform \texttt{KOBEsim}, $P_{\rm peak}$.}
 \label{fig:RVsp}
\end{figure}

\begin{figure}
  \begin{center}
     \subfigure{\includegraphics[width=93mm]{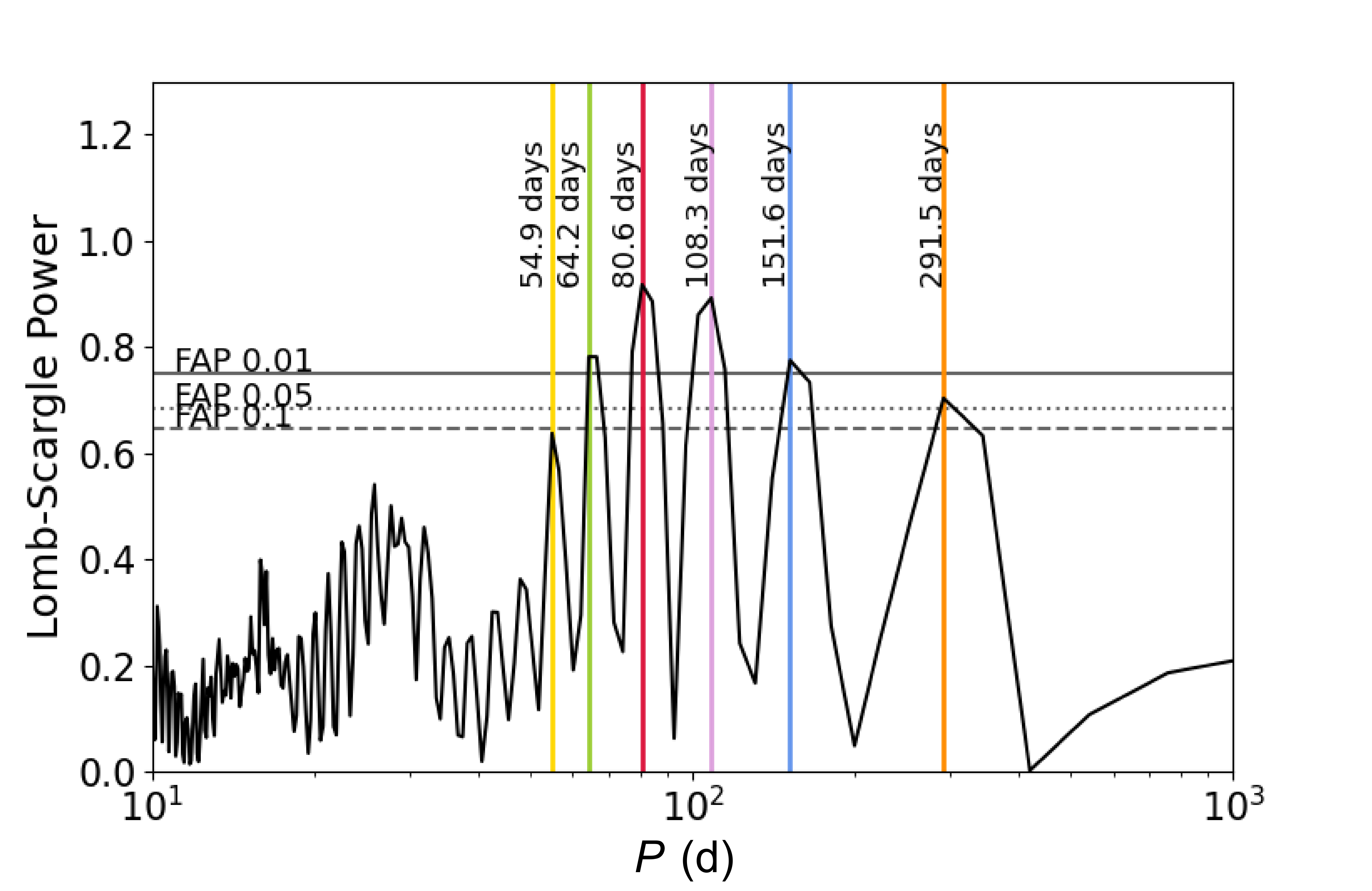}}
  \end{center}
 \caption{Periodogram for the studying case of $P_{\rm planet}$\,=\,107.57 d and $P_{\rm peak}$\,=\,81.47 d, with data gathered at the time of achieving the threshold $\ln(B_{10})$ > 6.}
 \label{fig:Periodogramsp}
\end{figure}

  To test this, we now focus on a given periodicity that maximizes the GLS periodogram, $P_{\rm peak}$, and we base our strategy on it. However, there is a planet orbiting the star at a different period, $P_{\rm planet}$. To illustrate this scenario, we start by simulating the first ten datapoints as a signal with the spurious period ($P_{\rm peak}$), constraining it to be compatible within 1$\sigma$ with the RV that would be induced by a planet at the true period ($P_{\rm planet}$). In this way we ensure that  the preliminary GLS periodogram proposes $P_{\rm peak}$. The next observations are generated using the true period, thus the GLS periodogram will start showing $P_{\rm planet}$ instead of    $P_{\rm peak}$. We perform this test for four random periods and the corresponding planetary masses that induce the same RV semi-amplitude in all cases ($K$ = 6 m\,s$^{-1}$ in this example). In Fig. \ref{fig:Psp} we show the evolution of the Bayesian evidence. The solid lines indicate the cases in which \texttt{KOBEsim} is targeting the correct orbital period. Also in Fig. \ref{fig:Psp}, we show a toy model for the active star case. We consider that after the first ten observations simulated with the spurious period $P_{\rm peak}$, the activity ceases, and thus the next datapoints follow a constant model ($K$\,=\,0 case, corresponding to a RV\,=\,V$_{\mathrm{sys}}$\,+\,E$_j$).

  From this simulation, we find that when targeting false periods with \texttt{KOBEsim}, the Bayes factor can be reduced in contrast with a successful case (when targeting a correct period all the candidate dates increase the Bayes factor, or are at least   compatible with no variation; see Sect. \ref{sec:simuldaily} and Fig. \ref{fig:KOBEsim_simul}). Specifically, in the case where we test a signal produced by stellar activity that disappears ($K$\,=\,0), it is quickly visible that the period is incorrect. Surprisingly, some spurious periods reach the planet detection according to the $\ln(B_{10})$ > 6 criterion. From the periodogram, we check that generally even selecting the optimum observing dates for a wrong period, the RV curve is well sampled and the GLS periodogram reveals the correct period (higher power for the true period than the targeted). Nonetheless, we find particular cases where this does not occur. As the semi-amplitude of this example is considerably high, the planet detection is reached too soon (with few datapoints) to sample the whole RV curve if $P_{\rm peak}$\,<\,$P_{\rm planet}$. To illustrate this situation, Fig. \ref{fig:RVsp} shows the RV curves in phase for the case of $P_{\rm peak}$\,=\,81.47\,d and $P_{\rm planet}$\,=\,107.57\,d. In the top panel  of Fig. \ref{fig:RVsp} we show the correct RV phase-folded curve (i.e., phase-folded with the true periodicity). In view of this curve, there are not enough data to claim a detection since the orbital phase has not yet been covered. When looking at the periodogram shown in Fig. \ref{fig:Periodogramsp}, we find   that there is a whole family of periods.

  In practice, this weak point can be easily circumvented by  looking at the periodogram daily. \texttt{KOBEsim} is the strategy to decide when to observe, but in any case it can replace our analysis. The user must be responsible for testing all the periods appearing in the GLS periodogram to check which is most favorable (e.g.,  calculating the Bayes factor for those competing planet hypotheses at different periods instead of using the null hypothesis). For this reason, the period chosen to be targeted with our strategy is not immutable, instead it must be updated as we gather additional data, as we indicate in the workflow scheme in Fig. \ref{fig:diagram}.

\subsection{Application to real data}
\label{sec:real}

\subsubsection{KOBE target}
\begin{figure*}
  \begin{center}
     \subfigure{\includegraphics[width=180mm]{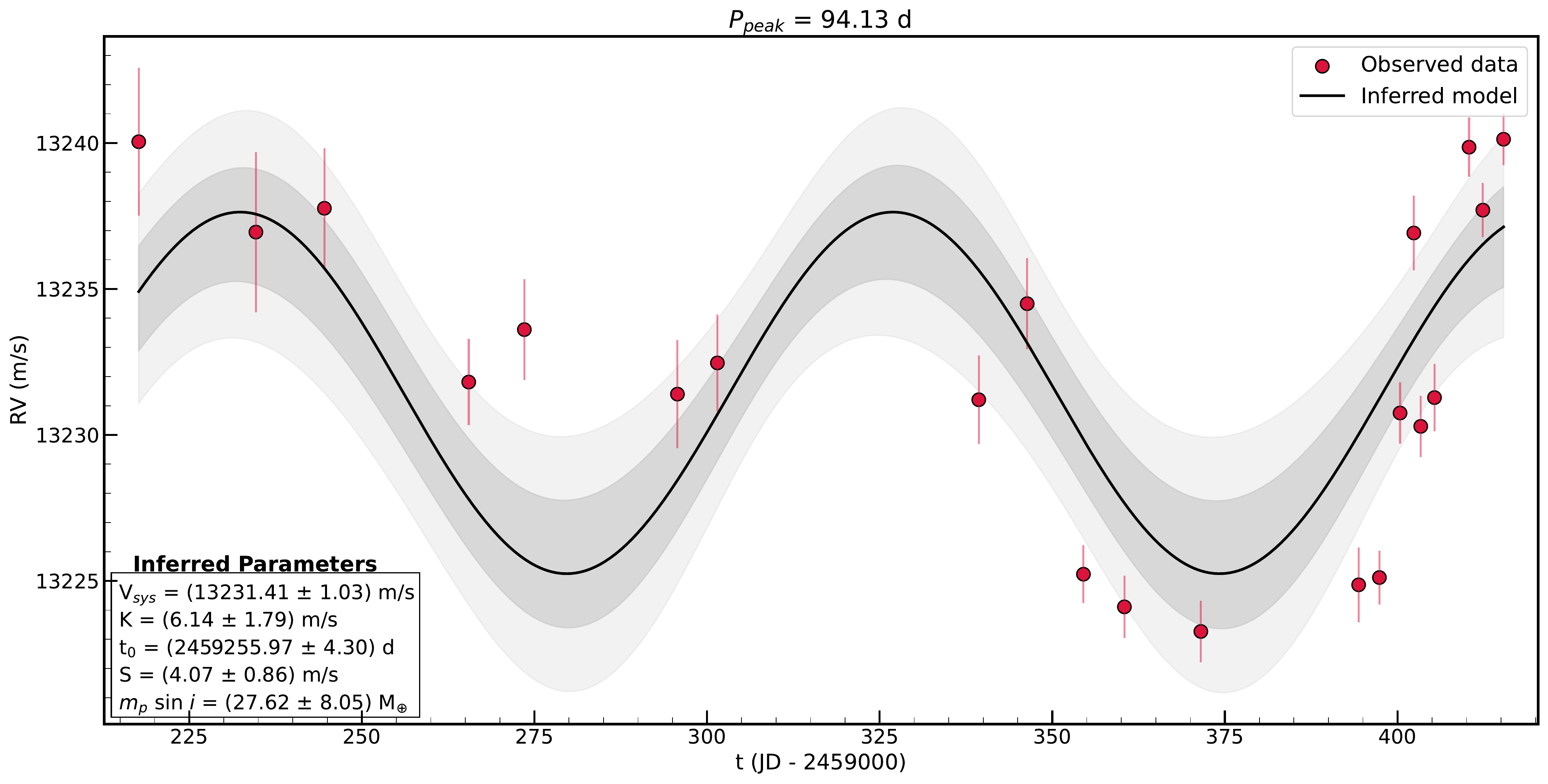}}
  \end{center}
 \caption{RV observed data (red dots) vs. time. The inferred model is shown as a solid line, using as prior $P_{\rm peak}$\,=\,94.13 d. The parameters resulting from the MCMC inference, as well as the corresponding mass lower limit, are collected in the  bottom left box. The shaded region shows the confidence interval at 1$\sigma$ (dark gray) and 2$\sigma$ (light gray) for the hypothesis with a planet.}
 \label{fig:RVreal}
\end{figure*}

We use the 21 first observed measurements from a particular KOBE target as a test bench. At this point a periodicity is visible inside the HZ ($P_{\rm peak}$\,=\,94.13 d). By taking this periodicity as our prior knowledge to run \texttt{KOBEsim} (i.e., assuming the RV is induced by an orbiting planet at $P_{\rm peak}$), we infer the parameters $\Theta_1$ (see Eq. \ref{eq:theta1}). In Fig. \ref{fig:RVreal}, we show  the RV time series (red dots), and the inferred model (solid line). With these inferred parameters, the lower limit for the planetary mass is $m_p$\,sin $i$\,=\,(27.62\,$\pm$\,8.05)\,$M_{\oplus}$.

\begin{figure}[h!]
  \begin{center}
    \subfigure{\includegraphics[width=85mm]{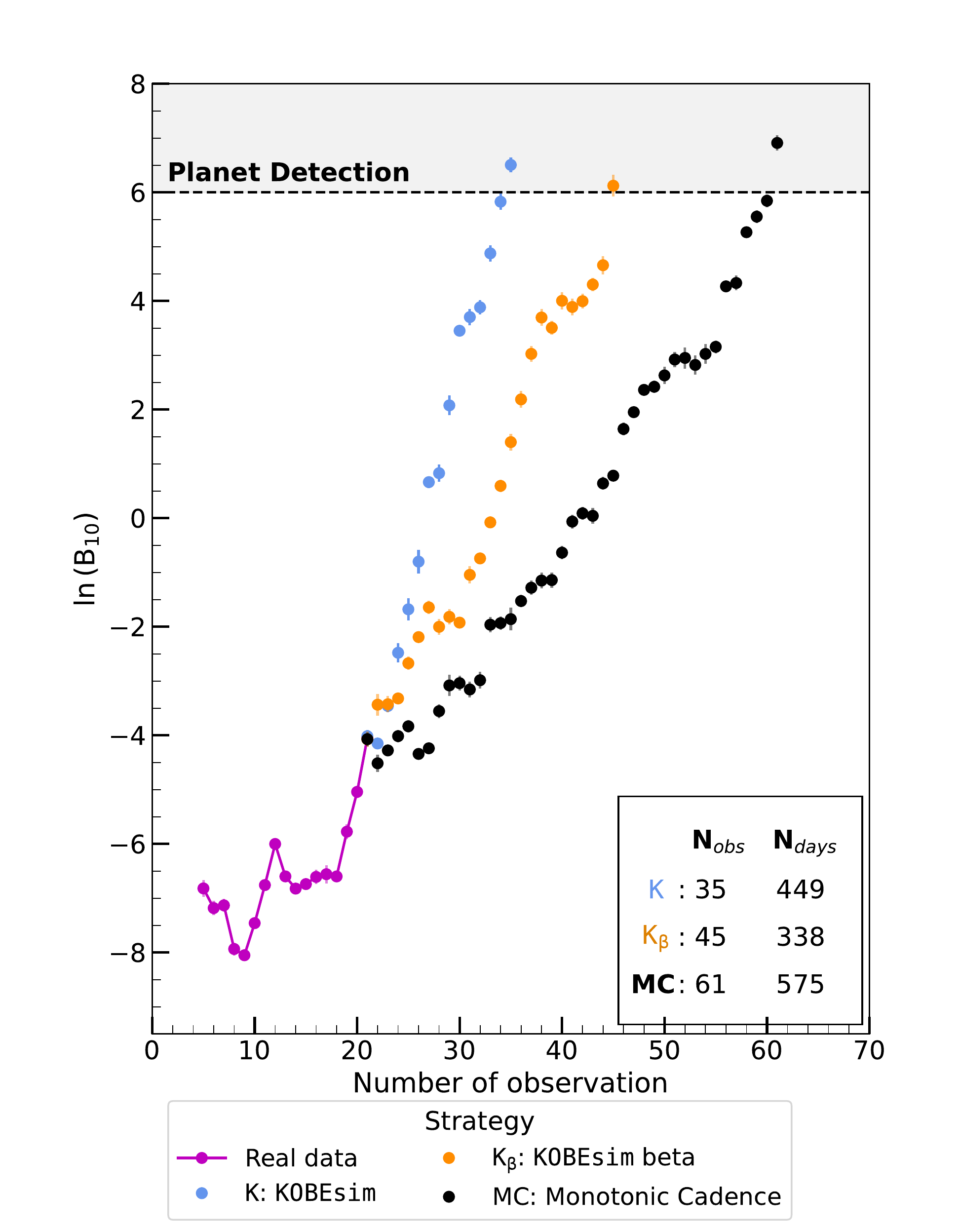}}
  \end{center}
 \caption{Prediction in the evolution of the logarithm of the Bayes factor in a real case of the KOBE experiment, assuming a planet at $P_{\rm peak}$\,=\,94.13 d. The number of observations and the time it   takes to detect the planet are compared using three different strategies: \texttt{KOBEsim} (K), \texttt{KOBEsim} beta (K$_{\beta}$), and spacing the observations in a fixed cadence (MC) of 6 d. Planet detection criterion: $\ln(B_{10})$ > 6.}
 \label{fig:seqreal}
\end{figure}

 We subsequently use these parameters to simulate the expected future RVs by making use of Eq. (\ref{eq:vj}). In such a way, after the 21st observation we estimate the evolution of $\ln{(B_{10})}$ and we compare the prediction for the three different observing strategies that we   consider  in the efficiency test explained in Sect. \ref{sec:eff} (i.e., \texttt{K}, \texttt{K$_{\beta}$}, and MC). This can be seen in Fig. \ref{fig:seqreal}, where the $\ln{(B_{10})}$ obtained from the observed data is shown (magenta line).

Based on the results of this analysis, and as long as the same trend on the RV data continues, we can detect a planet within the HZ of this star in less than a year. The expected improvement in the case of purely following the \texttt{K$_{\beta}$} instead of a MC is higher than 26\,\% in terms of the number of observations, and around 41\,\% in terms of time span. This improvement means, if this hypothetical planet actually exists in the system, that we could detect it a year earlier thanks to this approach.

\subsubsection{HD 102365}
\label{sec:realHD}

We take the RV data measured by \citet{2011ApJ...727..103T} to test the algorithm with a target from a different program and instrument. The goal is to evaluate the time that would have been saved if our strategy had been applied. We chose the system \object{HD~102365} since it has a relatively low-mass ($m_p\sin i$\,=\,16\,$M_\oplus$) isolated planet at a large period ($P$\,=\,122.1\,d), inducing a RV semi-amplitude of $\sim$ 3 m\,s$^{-1}$ while orbiting a G-dwarf star. Thus, it required a vast number of observations over a long time span to be detected. Additionally, the star is known to be chromospherically inactive with a value of -4.99 for $R^{\prime}_{HK}$ (\citealt{2017A&A...607A.124M}; \citealt{2018A&A...616A.108B}), which corresponds to an induced RV semi-amplitude of $\sim$\,41 cm\,s$^{-1}$ (\citealt{2017MNRAS.468.4772S}) and has a very slow rotation velocity ($\mathrm{v}$\,$\sin{i}$\,=\,0.7\,km\,s$^{-1}$), thus in this case activity does not play a relevant role. These observations were done with the UCLES \'echelle spectrograph (\citealt{1992ESOC...40..267D}) as part of the Anglo-Australian Planet Search (AAPS) program (\citealt{2001ApJ...551..507T}; \citealt{2020MNRAS.492..377W}).

We started testing the \texttt{K$_{\beta}$} strategy with the first ten real datapoints. The optimum date was chosen according to the algorithm, which simulates the RV at 20 sub-phases from the parameter inference of this reduced sample. In this case we considered that all the nights were available.

Once the date was decided, a new RV datapoint was added using the parameter inference from the whole sample (149 datapoints). To this new observation, we applied a Gaussian noise of mean and standard deviation of the median and standard deviation of the residuals. We took the same associated RV uncertainty as the corresponding real datapoint for the sake of being more fair with the comparison. Then, the process was repeated until  the planet was detected according to the Bayes factor. The target period, $P_{\rm peak}$, was updated at the beginning of each iteration by means of the GLS periodogram.

\begin{figure*}
  \begin{center}
     \subfigure{\includegraphics[width=180mm]{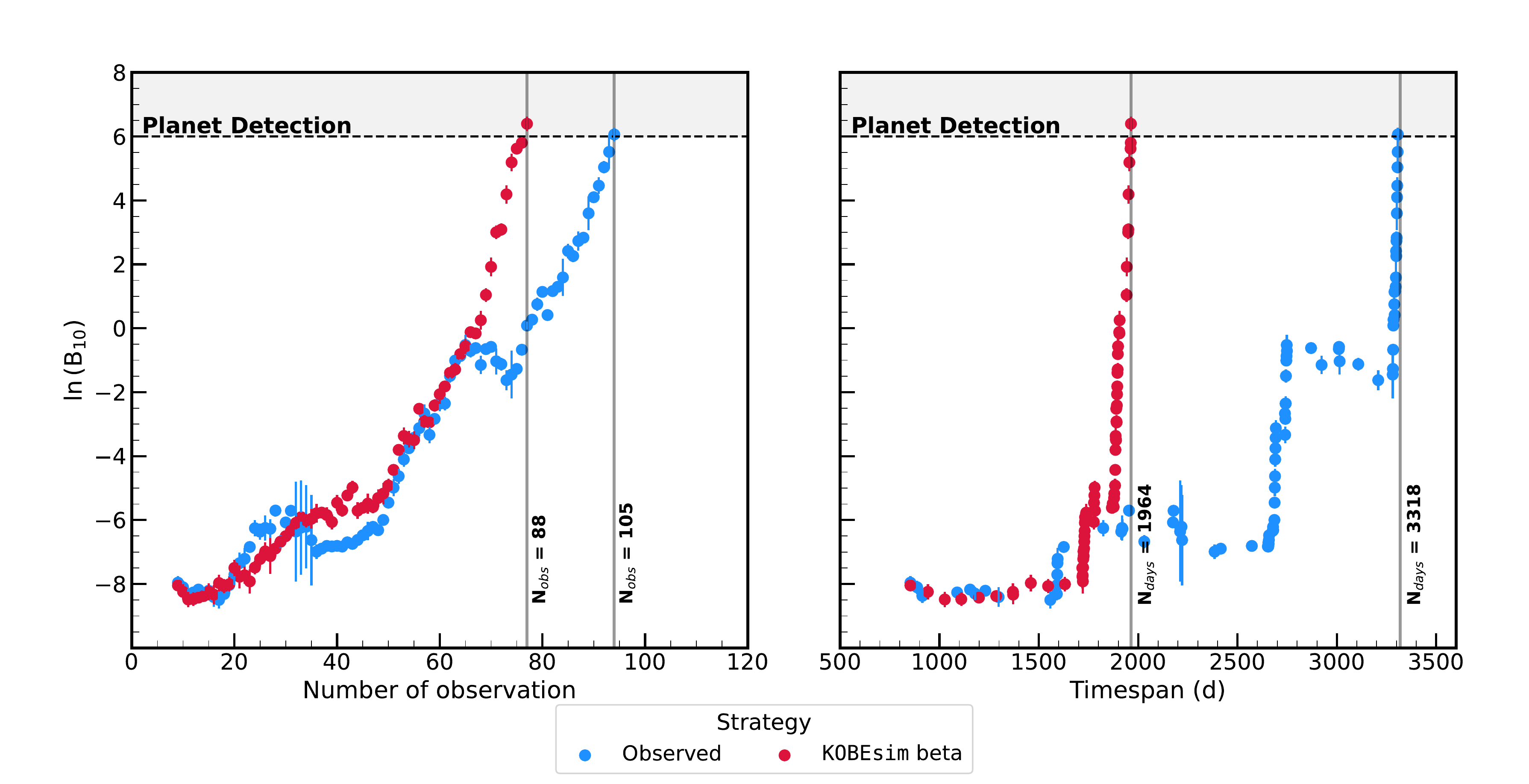}}
     \subfigure{\includegraphics[width=165 mm]{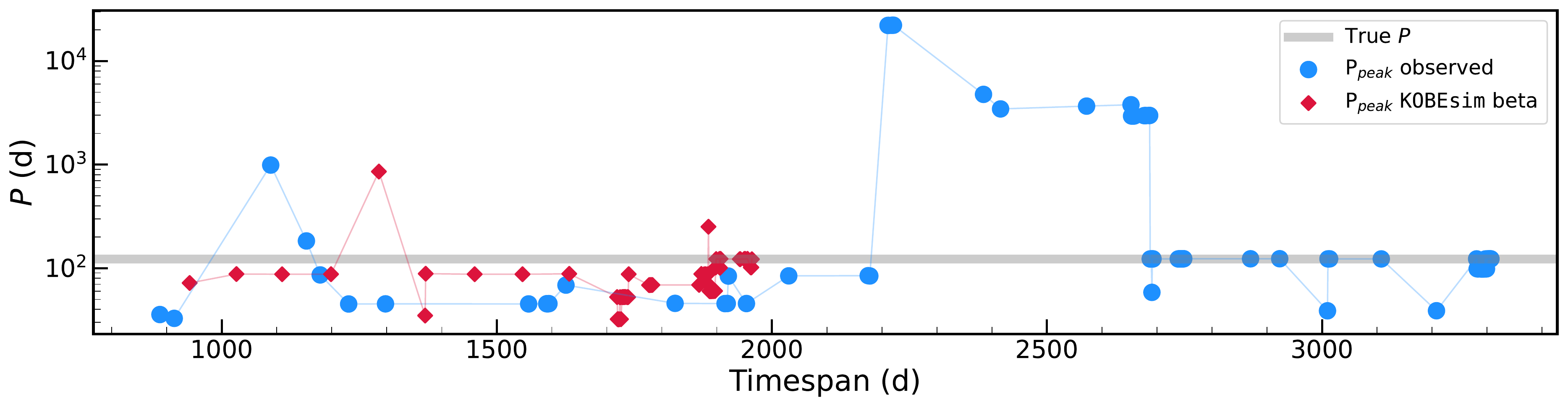}}
  \end{center}
 \caption{\texttt{KOBEsim} test with real \object{HD~102365} RV data. \textit{Upper panel}: Number of measurements (\textit{left}) and time span (\textit{right}) required to detect the planet comparing the real observations and the simulation for the \texttt{KOBEsim} beta strategy. \textit{Lower panel}: $P_{peak}$ evolution for both strategies.}
 \label{fig:HD102364}
\end{figure*}

In the upper panel of Fig. \ref{fig:HD102364} we show the evolution of the Bayes factor as a function of number of observations and time span, comparing our strategy with the real observations. We find that the number of measurements would be reduced by around  16\,\% (from 105 to 88 observations), and the time span by $\sim$41\,\% (from 3318 to 1964\,d). Since the exposure time varies from 200 to 400\,s, 1\,-- 2\,hours of telescope time would be   saved. In the lower panel of Fig. \ref{fig:HD102364} we display the evolution of the maximum period in the GLS periodogram for both cases. It is remarkable that it converges more quickly toward the true period of the planet using \texttt{K$_{\beta}$}.

\section{Conclusions}
\label{sec:concl}

In this paper we present \texttt{KOBEsim}, an algorithm designed to improve the efficiency in the process of gathering new RV measurements in blind-search surveys. This new observational strategy is developed aspiring to maximize the chances of success of the KOBE experiment (\citealt{2022A&A...667A.102L}). It is a Bayesian approach using the Bayes factor as a metric to measure how well supported  the planet hypothesis is at a given orbital period ($H_1$) compared to the null hypothesis where there is no planet in the system ($H_0$). Given the previous RV data and an orbital period to target, \texttt{KOBEsim} proposes a priority calendar to observe the star again according to the expected increase in this quantity. By weighting the increment in the Bayes factor with a beta distribution, we find a trade-off between number of measurements and time span necessary to claim a planet detection. We demonstrate its effectiveness in speeding up the planet detection when stellar activity is well characterized, being especially useful in lighter planets for which the improvement is nearly  50\,$\%$ in comparison with monotonic cadence strategies. This improvement can be decisive to detect rocky planets within the HZ in reasonable time spans.

These results show the importance of a continuous monitoring of the measurements in RV studies of planet searches and demonstrates that simple monotonic cadence strategies are not efficient, wasting more telescope time than required to confirm or characterize a given planet. It should be  noted that, even though designed for blind-search surveys, it   can also be highly useful in the follow-up of transiting candidates since the orbital period is clear. The approach described in this paper can be easily implemented for any instrument. High-resolution spectrographs able to detect planetary signals and installed in observatories offering service mode observations (e.g., ESO/Paranal or CAHA) should have the capability to adapt the observing strategy to improve their efficiency, which nowadays is difficult to achieve due to how observing time of the different programs is allocated. Offering the user the flexibility to adapt the cadence of the observations on a daily basis is a   benefit to  the community and to the observatories. In this regard, GTO programs can be highly benefited as they enjoy  wider freedom in their schedules; it is an opportunity to save time and to favor the detection of the most elusive planets for the upcoming generation of instruments, such as HARPS3 (\citealt{2016SPIE.9908E..6FT}) or NIRPS (\citealt{2017SPIE10400E..18W}).

Throughout the manuscript we have mentioned some caveats to bear in mind when using \texttt{KOBEsim}. First of all, the user has to take care of the preprocessing of the data in order to mitigate any RV signal induced by activity. The RV time series should be corrected from activity before running \texttt{KOBEsim}, for instance subtracting linear and quadratic trends. The use of activity indicators such as line-bisector or the chromospheric contribution of the H and K Ca lines are also useful to disentangle the planetary and activity component in the signal (e.g., \citealt{2001A&A...379..279Q}). Second, as concluded in Sect. \ref{sec:false}, the period to be targeted must always be updated   after a new observation is added to the dataset. Third, the implementation in the code of a multi-planetary system model (hypothesis $H_{n\,planets}$) is straightforward. We are conscious of the scientific value of this utility;  \texttt{KOBEsim} is not only   a tool to boost planet detection in single planet systems or in multi-planetary systems where no planets have been yet detected, but it can enable us to determine more quickly whether there is more than one planet inducing the signal. In the same regard, it would also be   interesting not to always compare the evidence of the model with the null hypothesis, but with another planet hypothesis orbiting at other period to deal with aliasing. These implementations would make our algorithm more powerful, but are yet to be tested and are beyond   the scope of the present work.

\begin{acknowledgements}
    We are grateful to Dr.\,Rodrigo Díaz for his useful suggestions as referee helping to improve the quality of the manuscript. O.\,B.\,-R., J.\,L.\,-B. and A.\,C.\,-G. acknowledge financial support received from "la Caixa" Foundation (ID 110000434) and from the European Unions Horizon 2020 research and innovation programme under the Marie Sklodowska-Curie grant agreement No 847648, with fellowship code LCF/BQ/PI20/11760023. This research has also been partly funded by the Spanish State Research Agency (AEI) Projects No.PID2019-107061GB-C61 and No. MDM-2017-0737 Unidad de Excelencia "Mar\'ia de Maeztu"- Centro de Astrobiolog\'ia (INTA-CSIC).
This work was supported by Fundação para a Ciência e a Tecnologia (FCT) and Fundo Europeu de Desenvolvimento Regional (FEDER) via COMPETE2020 - Programa Operacional Competitividade  e  Inter-nacionalização by  these grants: UIDB/04434/2020; UIDP/04434/2020; PTDC/FIS-AST/32113/2017   \& POCI-01-0145-FEDER-032113; PTDC/FIS-AST/28953/2017    \& POCI-01-0145-FEDER-028953
A.\,M.\,S. acknowledges support from FCT through the Fellowship 2020.05387.BD. and POCH/FSE (EC).
    O.\,D.\,S.\,D. is supported in the form of work contract (DL 57/2016/CP1364/CT0004) funded by FCT.
    J.\,P.\,F. is supported in the form of a work contract funded by national funds through FCT with reference DL 57/2016/CP1364/CT0005.
    A.\,M.\,S. acknowledges financial support from the French Programme National de Planétologie (PNP, INSU).
\end{acknowledgements}

\bibliography{references}

\begin{appendix}

\section{\texttt{KOBEsim} inputs and outputs}
\label{sec:A}

\noindent There are 12 fields accepted as input; they are all found in Table \ref{tab:inputs}. The observatory coordinates, the star name, and the file containing the RV time series are mandatory. As the orbital period is an input of the code, we highly recommend that the user   perform a careful study of the period to be targeted, for instance using the $\ell$1 periodogram (\citealt{2017MNRAS.464.1220H}) as a complementary method to the GLS periodogram.

As a result of running \texttt{KOBEsim}, a prioritized list of calendar dates is delivered in the form of an ascii file in csv format. Each row corresponds to a candidate future observing date ranked by preference: from highest to lowest weighted $\Delta\ln(B_{10})$. The columns from left to right are calendar date (format year-month-day), JD, the corresponding orbital phase, the expected $\ln(B_{10})$ and its associated uncertainty, and the increase in $\Delta\ln(B_{10})$ and its uncertainty. An example of this output file is shown in Table \ref{tab:csv1} for the simulated case of Sect. \ref{sec:simuldaily}. For a more illustrative inspection of the results, \texttt{KOBEsim} also returns a plot of $\ln{(B_{10, n+1})}$ versus the orbital phase. See Sect. \ref{sec:simuldaily} for details.

\setlength{\tabcolsep}{1.5pt}
\begin{table*}[h]
    \caption[]{Inputs of the \texttt{KOBEsim} code: \texttt{obs} (or \texttt{obs$\_$n}), \texttt{star}, and \texttt{file} are mandatory.}
    \label{tab:inputs}
    \centering
    \begin{tabular}{@{}cccc@{}}
    \hline \hline
    Symbol         & Parameter         &    Default      & Description \\ \hline
    \texttt{obs}   & Observatory coordinates       &    -    & Observatory coordinates (deg) and altitude (m) \\
    \texttt{obs$\_$n}   & Observatory name      &    -    & Observatory name to obtain the coordinates \\
    \texttt{star}  & Star name         & -       & To search for its coordinates with SIMBAD (\citealt{2019AJ....157...98G})  \\
    \texttt{file}  &   RV data file       &  -      & FITS or ascii format with columns: JD, RV and $\Delta$RV (m\,s$^{-1}$)\\
    \texttt{sch}   &   Schedule        &  \texttt{None}  & Ascii file with the program asigned JD. If \texttt{None}, all dates are considered\\
    \texttt{P}     &   Period          &  GLS periodogram peak     & Proposed period (d) to be targeted with \texttt{KOBEsim} \\
    \texttt{t$_0$}   &   Inferior conjunction time &  \texttt{None}               & If provided, it is used as prior in the MCMC inference\\
    \texttt{minalt}&   Minimum altitude& 20 deg          & Minimum altitude (deg) constraint to observe the target\\
    \texttt{texp}  &   Exposure time   & 700 s           & Exposure time (s) for the target\\
    \texttt{Nph}   &   Number of sub-phases    &    20     & Number of splits of the orbital phase ($N_{\rm phases}$)\\
    \texttt{beta}  &   Beta distribution &    \texttt{True}& To reduce the time between observations through a beta distribution\\
    \texttt{ab}    &   Beta parameters &  a = 1, b = 5 & Parameters of the beta distribution, $\beta$(a, b)\\
    \texttt{wh}    &   Pre-whitening    &   \texttt{False} & To substract a linear and a quadratic contribution in the RV data\\
    \texttt{max$\_$da}&   Maximum days apart    &   90\,d & Maximum days apart to search for the next optimum observing date\\
    \texttt{n}    &   Number of steps    &   20\,000 & Number of steps per walker for the emcee warm-up phase\\
    \texttt{nw}    &   Multiple number of walkers    &   4  & Multiple of the number of parameters for the number of walkers\\
    \hline
    \end{tabular}
   \end{table*}

\begin{table*}[]
\caption[]{Output csv file of \texttt{KOBEsim} for the testing target with simulated data.}
\label{tab:csv1}
\centering
\begin{tabular}{c c c c c c c}
\hline\hline

\textbf{Calendar\_day} & \textbf{JD} &  \textbf{phase} & \textbf{lBF} & \textbf{sigma\_lBF} & \textbf{delta\_lBF} & \textbf{sigma\_delta\_lBF} \\
\hline
2021-04-06             & 2459311     & 0.170          & -3.393       & 0.028               & 0.348               & 0.055                      \\
2021-04-16             & 2459321     & 0.339          & -3.157       & 0.037               & 0.584               & 0.061                      \\
2021-04-17             & 2459322     & 0.356          & -3.200       & 0.043               & 0.541               & 0.064                      \\
2021-04-03             & 2459308     & 0.119          & -3.520       & 0.039               & 0.221               & 0.062                      \\
2021-03-28             & 2459302     & 0.107          & -3.578       & 0.038               & 0.163               & 0.061                      \\
2021-04-20             & 2459325     & 0.407          & -3.249       & 0.037               & 0.492               & 0.061                      \\
2021-03-31             & 2459305     & 0.068          & -3.576       & 0.030               & 0.166               & 0.057                      \\
2021-05-04             & 2459339     & 0.645          & -3.567       & 0.077               & 0.174               & 0.091                      \\
2021-05-08             & 2459343     & 0.712          & -3.503       & 0.049               & 0.238               & 0.069                      \\
2021-05-05             & 2459340     & 0.662          & -3.566       & 0.036               & 0.175               & 0.060                      \\
2021-05-17             & 2459352     & 0.865          & -3.413       & 0.093               & 0.328               & 0.105                      \\
2021-05-20             & 2459355     & 0.916          & -3.506       & 0.043               & 0.235               & 0.065                      \\
2021-05-23             & 2459358     & 0.967          & -3.616       & 0.046               & 0.125               & 0.066                      \\
2021-06-06             & 2459372     & 0.204          & -3.297       & 0.035               & 0.444               & 0.060                      \\
2021-06-21             & 2459387     & 0.459          & -3.527       & 0.057               & 0.214               & 0.075                      \\
2021-06-24             & 2459390     & 0.510          & -3.682       & 0.050               & 0.059               & 0.069                      \\

\hline
\end{tabular}
\end{table*}
\end{appendix}

\end{document}